\documentclass{article}

\usepackage{graphicx}

\usepackage{amssymb}
\usepackage{amsbsy}

\begin{document}


\title{Optimal Physical Multipoles}


\author{Charles Baker \\
	Departments of Physics, Biology and Mathematics \\
	Virginia Tech, Blacksburg, VA \\
	cbaker@vt.edu
	\and
	Ramu Anandakrishnan \\
	Departments of Computer Science and Physics \\
	Virginia Tech, Blacksburg, VA \\
	ramu@cs.vt.edu
	\and
	Alexey Onufriev \\
	Departments of Computer Science and Physics \\
	Virginia Tech, Blacksburg, VA \\
	alexey@cs.vt.edu \\
}

\date{\today}

\maketitle

\begin{abstract}

Point multipole expansions are widely used to gain physical
insight into complex distributions
of charges 
and to reduce the cost of computing interactions
between such distributions.
However, practical applications that typically retain only a few leading terms
may suffer from unacceptable loss of accuracy in the near-field.
We propose an alternative approach
for approximating electrostatic charge distributions,
Optimal Physical Multipoles (OPMs),
which optimally represent the original charge distribution
with a set of point charges. 
By construction, approximation of electrostatic potential based on OPMs retains 
many of the useful 
properties of the corresponding point multipole expansion,
including the same asymptotic behavior of the approximate potential 
for a given multipole order. At the same time, OPMs 
can be significantly more accurate in the near field:
up to 5 times more accurate for some of the charge distributions
tested here which are relevant to biomolecular modeling.
Unlike point multipoles, for point charge distributions
the OPM always converges to the original
point charge distribution at finite order. 
Furthermore, OPMs may be more computationally efficient and
easier to implement into existing molecular simulations software packages
than approximation schemes based on point multipoles. 
In addition to providing
a general framework for computing OPMs to any order, closed-form 
expressions for the lowest order OPMs (monopole and dipole) are derived. 
Thus, for some practical applications Optimal Physical Multipoles 
may represent a preferable alternative to point multipoles.

\end{abstract}






\section{Introduction}

Point multipole expansions are widely used
to gain physical insight by providing a simplified expression
for a complex distribution of sources of potential fields,
such as electrostatic potential due to a charge distribution.
Point multipoles provide a means of decoupling the underlying features
of a source distribution from the observation point.
Thus, one can obtain physically meaningful insights into the
macroscopic properties of the distribution,
such as the familiar dipole moment.

In addition to having theoretical utility, the point multipole
expansion has been used to simplify practical calculations. 
For example, algorithms such as the fast multipole method \cite{Greengard1987}, use point 
multipoles to reduce the computational complexity of calculating pairwise
interactions between large charge distributions. 
Let $R_{0}$ be the distance of the furthest charge in the distribution 
from the expansion center. Since each successive term
in the multipole expansion decays more rapidly with distance than the previous
term, at large distances $R \gg R_{0}$ the series converges quickly and so these methods
are able to obtain reasonably accurate results by keeping only 
the lowest order terms.
However, at distances not much larger than $R_{0}$,
the accuracy of the 
approximation deteriorates quickly. Since, in practice,
the potential often needs to be calculated in regions where 
the assumption $R \gg R_{0}$ does not hold
(the near-field), the point multipole
expansion is suboptimal for many practical calculations. 
For example, in atomistic molecular simulations, amino acids
interacting inside a single protein 
are rarely more than several times $R_0$ apart.
For these calculations, point multipoles can be expected to provide a
suboptimal approximation of the original distribution. 

\begin{figure}[h!]
\centering
\includegraphics[width=10cm]{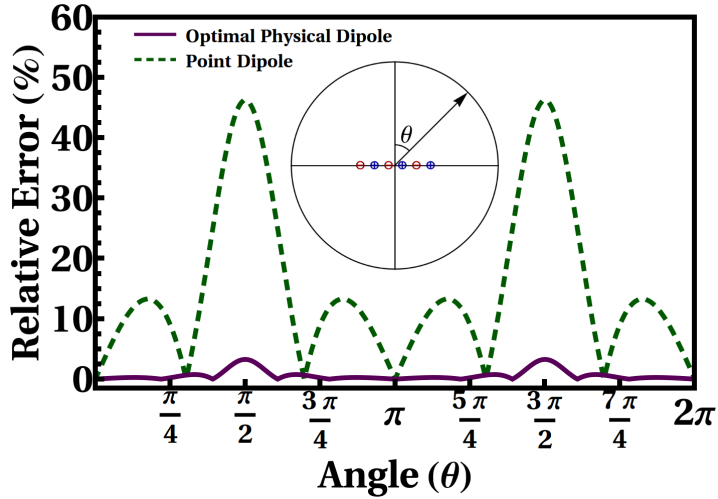}
\caption{\label{fig:Motivation}
Relative errors 
in electrostatic potential around
an example net neutral charge distribution consisting 
of 6 charges with equal magnitude spaced equidistant on a line and alternating 
negative to positive, see inset. 
An Optimal Physical Dipole produces
less than 3.5\% error for all angles (solid purple line),
whereas an optimally placed point dipole produces 
over 45\% error at some angles (dashed green line).
The Optimal Physical Dipole consists of a negative
and positive charge, of equal magnitude, placed on the left and 
right ends of the original charge
distribution respectively. On the other hand, an optimally placed point dipole,
as defined in \cite{Gramada2008,Platt1996}, is located at the center 
of geometry. 
The relative error is computed as 
$\left(\left|\Phi(2 R_{0},\theta)-\Phi_{ref}(2 R_{0},\theta)\right|\Bigl/\sqrt{(1/4\pi) \intop_{0}^{2\pi}\intop_{0}^{\pi}(\Phi_{ref}(2 R_{0},\theta))^{2} \sin(\theta) d\phi d\theta)}\right) \times 100 \% $
where $\Phi(2 R_{0},\theta),\Phi_{ref}(2 R_{0}, \theta)$ are the electrostatic 
potentials of the approximate and the reference (original) charge 
distributions respectively. 
The electrostatic potentials are computed at a distance 
$2 R_{0}$ from the center of geometry, where $2 R_{0}$ is
the size of the charge distribution (twice the distance from the center 
of geometry to the outermost charge in the original distribution). 
}
\end{figure}

We investigate an alternative to the point multipole expansion
for approximating charge distributions
-- Optimal Physical Multipoles. Optimal Physical Multipoles approximate a charge distribution by
a small number of point charges spatially distributed in an ``optimal'' way, 
to be precisely defined below. 
Since OPMs have a finite size,
they may provide better  
representation of the original spatially extended 
charge distribution, i.e. a more accurate representation of the potential in the near-field. 
Consider for example the extreme case where the charge distribution consists
of 6 charges of equal magnitude spaced equidistant on a line and alternating 
negative to positive (Figure \ref{fig:Motivation}).
For such a distribution, 
an optimally placed point dipole still produces more than 25 times the average 
error produced by an Optimal Physical Dipole at a distance of
$2R_{0}$ from the center of geometry of the original charge distribution.
Furthermore, later we will prove that 
for at least the lowest order OPMs,
i.e. Optimal Physical Monopole and Optimal Physical Dipole, OPMs
are at least as accurate as an 
equivalent order point multipole. From a practical standpoint, 
Optimal Physical Multipoles may be 
easier to implement in applications that
already utilize point charges, i.e. in many molecular dynamics packages
\cite{Pearlman1995,schlick2010molecular} 
and are potentially more computationally efficient than point multipoles.

Representing charge distributions by a small number of point charges
is not by itself a novel idea.
There are a number of methods, such as RESP \cite{Bayly1993}, CHELP \cite{Chirlian1987},
CHELPG \cite{Breneman1990}, CHELMO \cite{Sigfridsson1998},
coarse graining \cite{Basdevant2007,Izvekov2005,Anandakrishnan2010} and others \cite{Swart2001}
that empirically fit a set of point charges to a given charge distribution
by minimizing various error metrics in electrostatic potential over some
volume or surface surrounding the charge distribution.
A key difference between these methods and Optimal Physical Multipoles 
introduced here is that OPMs inherit the physically appealing 
asymptotic properties of the point multipole approximation,
i.e. the error in potential falls off at least as fast as $1/R^{k+1}$,
where $R$ is the distance from the origin and
$k$ is the highest order of the multipole terms retained in the expansion.

The rest of this work is organized as follows. We 
first review the multipole expansion concept.
Next, we describe the theoretical basis for Optimal Physical Multipoles. We
then use this theoretical formalism to derive closed-form expressions
for the Optimal Physical Monopole and Dipole, discuss their properties
and compare their accuracy with the corresponding point multipoles.
Finally we provide a cost-benefit analysis of Optimal Physical Multipoles 
versus the point multipole method and discuss the prospect
of expansions to higher orders. Potential uses and future work is discussed 
in ``Conclusions". Although point multipoles have 
been used to study a wide range of
potentials, this proof-of-concept study focuses on just one of the most common
application of the concept - electrostatics.
Nevertheless, we believe that the ideas presented here will be generally applicable
to other $1/R$, Coulomb-like, potentials as well.

\section{Multipole Expansion}

Here we will give a brief overview of the familiar formalism of the
point multipole expansion.
Since many practical applications, such as molecular dynamics simulations, 
use point charges, for notational simplicity we will 
consider discrete charge distributions, but the results 
also hold for continuous distributions. 
	
	Consider a set
of N point charges $q_{n}$ $(n=1,2,...,N)$ located at positions
$\boldsymbol{r_{n}}$ around some chosen origin. Then the potential
$\Phi(\boldsymbol{R})$, of this distribution at a point $\boldsymbol{R}$
from that origin is given by the familiar Coulomb Potential
\begin{equation}
  \Phi(\boldsymbol{R})=\frac{1}{4\pi\epsilon_{0}}
    \sum_{n=1}^{N}\frac{q_{n}}{||\boldsymbol{R}-\boldsymbol{r_{n}}||}
  \label{eq:Coulomb Potential}
\end{equation}
For distances $R>R_{0}$
where $R=||\boldsymbol{R}||$ and $R_{0}=max(||\boldsymbol{r_{n}}||)$, 
a Taylor series expansion of the potential above
gives the classic multipole expansion.
In Cartesian coordinates we obtain
\begin{eqnarray}
   \Phi(\boldsymbol{R}) & = & \frac{1}{4\pi\epsilon_{0}}
      \Biggl(\frac{1}{R} q +
      \frac{1}{R^{2}}\sum_{i=x,y,z}\hat{R_{i}}\boldsymbol{p_{i}} +
      \frac{1}{R^{3}}\sum_{i,j=x,y,z}\hat{R_{i}}\hat{R_{j}} Q_{ij} \nonumber \\ & &
      \qquad \quad + \frac{1}{6}\frac{1}{R^{4}}\sum_{i,j,k=x,y,z}\hat{R_{i}}\hat{R_{j}}\hat{R_{k}} O_{ijk} +
      \dots\Biggr)
   \label{eq:Cartesian_Expansion}
\end{eqnarray}
where 
\begin{eqnarray}
  q & = & \sum_{n=1}^{N}q_{n}\\
  \boldsymbol{p_{i}} & = & \sum_{n=1}^{N}q_{n}r_{n,i}\\
  Q_{i,j} & = & \frac{1}{2} \sum_{n=1}^{N}q_{n}\Bigl(3r_{n,i}r_{n,j}-(r_{n})^{2}\delta_{ij}\Bigr)\\
  O_{i,j,k} & = & \sum_{n=1}^{N}q_{n}\Bigl(15r_{n,i}r_{n,j}r_{n,k}-3(r_{n})^{2}(r_{n,i}\delta_{jk}+r_{n,j}\delta_{ik}+r_{n,k}\delta_{i,j})\Bigr)
  \label{eq:multipoles}
\end{eqnarray}
and $q,\boldsymbol{p},Q,O$
are known as the monopole, dipole, quadrupole and octupole moments
respectively, and $\delta_{ij}$ is the Kronecker delta.
The multipole moments are symmetric tensors where the lowest 
order non-vanishing multipole is origin independent. 

\section{Definition of Optimal Physical Multipoles}

For a given set of original charges $q_{i}$ $(i=1,2,...,N)$, we want
to determine a smaller representative set of charges $\bar{q}_{k}$
$(k=1,2,...,K < N)$
such that the potential due to these representative charges,
$\bar{\Phi}(\boldsymbol{R})$ best approximates the potential
of the original distribution, $\Phi(\boldsymbol{R})$.
In the following sections, we will outline the general method
for determining this optimal set of representative charges, i.e. 
the Optimal Physical Multipole representation. 

\subsection{The Error Metric}

Determining the best representative charge distribution
is critically contingent upon the definition of the error metric used.
In general we are concerned with obtaining the best representation
of the original potential at any arbitrary point in space
outside the distribution.
Thus, for the error metric, $\Delta$, one 
typically chooses the root mean square (RMS) of 
the error in potential over some volume $V$
excluding the volume $V_0$ containing the charge distribution
being approximated, i.e.
\begin{equation}
   \Delta^2=\frac{1}{V \notin V_0} \intop_{V \notin V_0} \left|\Phi(\boldsymbol{R})-\bar{\Phi}(\boldsymbol{R})\right|^2 dV
   \label{eq:RMS_ERROR}
\end{equation}
In principle, one can derive the optimal charge placement 
$\{\bar{q}_{n},\boldsymbol{\bar{r}_{n}}\}$ 
by minimizing the integral given in
Eq. (\ref{eq:RMS_ERROR}) with respect to the values of the 
new charges, $\{\bar{q}_{n}\}$ and
their positions $\{\boldsymbol{\bar{r}_{n}}\}$.
However, as the number of charges in the representative distribution
grows, this equation can be difficult to minimize, let alone to
find closed-form, analytic expressions for the placement and magnitude
of the charges composing the representative distribution.
Furthermore, charges chosen in this manner are not guaranteed to
have the same multipole moments 
as the original distribution \cite{Sigfridsson1998}.
This can lead to misinterpretation of the properties of the 
distribution and, potentially, to unphysical results. At the very least, 
we would like the new representation to inherit the
same transparent asymptotic behavior of the corresponding point multipole
expansion, but with greater accuracy expected from an extended distribution 
that can better mimic the original. 
To simplify the problem, we recast Eq. (\ref{eq:RMS_ERROR}) in spherical coordinates
and consider the error inside a spherical shell centered on the 
chosen multipole expansion center, and with arbitrary outer radius $\widetilde{R}>R_0$, 
where $R_0$ is defined as before, i.e. the distance from the 
expansion center to the outermost point charge. 
The error metric now becomes
\begin{equation}
   \Delta^2=\frac{3}{4 \pi (\widetilde{R}^3 - R_0^3)} \intop_{R_0}^{\widetilde{R}}\intop_{0}^{2\pi}\intop_{0}^{\pi}
      \left|\Phi(\boldsymbol{R})-\bar{\Phi}(\boldsymbol{R})\right|^2
      R^{2}\sin(\theta)d\theta d\phi dR
   \label{eq:RMS_ERROR_SPHERICAL}
\end{equation}
where $\theta$ and $\phi$ are the usual spherical coordinate
inclination and azimuth angles.

In spherical coordinates, the multipole expansion is given by
\begin{equation}
   \Phi(\boldsymbol{R}) = \frac{1}{\epsilon_{0}}
      \sum_{\ell=0}^{\infty} \sum_{m=-\ell}^{m=\ell}
      \frac{1}{2\ell+1} \frac{Y_{\ell}^{m}(\theta,\phi)}{R^{\ell+1}}
      q_{\ell}^{m}
   \label{eq:Spherical_Expansion}
\end{equation}
where $Y_{\ell}^{m}$ are the standard spherical harmonics, $*$ is
the complex conjugate,
$q_{\ell}^{m}$ are the spherical multipole moments, and
$\ell$ is the multipole order.
 \begin{equation}
 q_{\ell}^{m}=\sum_{n=1}^{N}q_{n}r_{n}^{\,\ell}Y_{\ell}^{m}(\theta_{n},\phi_{n})^{*}
 \label{eq:Spherical_Multipole}
 \end{equation}
Using this expansion as our error metric, Eq. (\ref{eq:RMS_ERROR_SPHERICAL}),
becomes
\begin{equation}
   \Delta^2=\frac{3}{4 \pi (\widetilde{R}^3 - R_0^3)}\intop_{R_{0}}^{\widetilde{R}}\intop_{0}^{2\pi}\intop_{0}^{\pi}
      \left|\frac{1}{\epsilon_{0}}\sum_{\ell=0}^{\infty}
      \sum_{m=-\ell}^{m=\ell}\frac{1}{2\ell+1}\,
      \frac{Y_{\ell}^{m}(\theta,\phi)}{R^{\ell+1}}
      (q_{\ell}^{m}-\overline{q_{\ell}^{m}})
      \right|^2 R^{2}\sin(\theta) d\theta d\phi dR
   \label{eq:MULTIPOLE_RMS_ERROR}
\end{equation}
where $q_{\ell}^{m}$ and $\overline{q_{\ell}^{m}}$
are the spherical moments of the original and representative charge
distributions respectively.
Since the spherical harmonic integral $\intop_{0}^{2\pi}\intop_{0}^{\pi}
   \left|Y_{\ell}^{m}(\theta,\phi)\right|^2 \sin(\theta) d\theta d\phi = 1$,
Eq. (\ref{eq:MULTIPOLE_RMS_ERROR}),
can be further simplified to the following form \cite{Platt1996},
\begin{equation}
   \Delta^2 = \frac{3}{4 \pi \epsilon_{0} (\widetilde{R}^3 - R_0^3)}\intop_{R_{0}}^{\widetilde{R}} \sum_{\ell=0}^{\infty}
      \frac{1}{\left(2\ell+1\right)^{2} R^{2(\ell+1)}}
      \sum_{m=-\ell}^{m=\ell}|q_{\ell}^{m}-\overline{q_{\ell}^{m}}|^{2} dR
   \label{eq:MULTIPOLE_RMS_ERROR_SIMP}
\end{equation}

From the structure of Eq. (\ref{eq:MULTIPOLE_RMS_ERROR_SIMP}) we see that
minimizing the difference
between the successive multipole moments of the original and Optimal Physical Multipole
distributions is equivalent to minimizing the total error,
i.e. by minimizing each term in the error expansion we minimize
the RMS error in electrostatic potential. 
Note that the procedure does not depend on the parameter $\widetilde{R}$,
and thus the method does not require explicit integration over
a given region. 
The use of the multipole expansion in this way 
allows for the sought after distinct separation of terms by the rate at which
they decrease as a function of $R$, i.e. the monopole term falls off as $\frac{1}{R}$,
the dipole falls off as $\frac{1}{R^{2}}$, etc.
A representation which makes terms up to 
order $k$ in Eq. (\ref{eq:MULTIPOLE_RMS_ERROR_SIMP}) equal to zero 
will produce total error whose leading term falls off as $1/R^{k+1}$. 
 
\subsection{Optimal Physical Multipole Approximation}

We define an Optimal Physical Multipole of order $k$ to be a set of 
charges satisfying the following conditions:
\begin{enumerate}

\item It is the smallest set of point charges which make every term in
the error expansion, Eq. (\ref{eq:MULTIPOLE_RMS_ERROR_SIMP}), through order $k$ equal to zero (for the given charge distribution being approximated).
\item If the charge magnitudes and positions are not uniquely determined by condition 1,
the charges are chosen such that the $k+1,k+2,...$ terms in
the error expansion are sequentially globally minimized until all the charges
are uniquely defined. 

\end{enumerate}

Optimal Physical Multipoles as defined above have the useful property that 
for any point charge distribution they will, at some order $k$,
reproduce the original point charge distribution.
For, at some order $k$, the 
OPM will require the same number of charges $N$ as the original distribution to
satisfy the first condition of the Optimal Physical Multipole definition. Furthermore, the OPM 
must be made identical to the original distribution to satisfy condition 2 from the definition above.
To see how this arises note that the error expansion contains terms of the form 
$(q_{\ell}^{m}-\overline{q_{\ell}^{m}})^{2}$. Thus, an Optimal Physical Multipole
which is identical to the original distribution will 
have $\overline{q_{\ell}^{m}}=q_{\ell}^{m}$ for all $\ell,m$
and clearly having $\overline{q_{\ell}^{m}}=q_{\ell}^{m}$ for all $m$
will be a global minimum of the error expansion term of order $\ell$. 
Since making the OPM identical to the original distribution will always provide a 
global minimum to all orders, setting the OPM equal to the original distribution provides 
the only way to place the charges such that they satisfy the second condition defining an OPM. 
We note that the convergence of Optimal Physical Multipoles to the original distribution after a finite
number of computations sets OPMs apart 
from point multipole expansions which 
generally require an infinite number of terms
to exactly reproduce a given charge distribution.  

The minimizations of the error metric in Eq. (\ref{eq:MULTIPOLE_RMS_ERROR_SIMP}) which
are required to define an Optimal Physical Multipole
can be done numerically to arbitrary order. This numeric procedure for calculating the Optimal Physical Multipole
representation may be particularly
useful in situations where the charge distributions are relatively static and thus the
optimal representation does not need to be recalculated at each time step
of a given simulation. Ideally, however, one would like analytic expressions
that can be used to compute OPMs at a reduced computational
cost.

\section{Analytic Expressions for Low Order Optimal Physical Multipoles}

In the following sections, we apply the general methodology developed in section 3,
to derive low order OPMs and test their accuracy.  
In sections 4.1 and 4.2, we obtain analytic forms for the lowest order Optimal Physical Multipoles
in two important special cases, namely,
we define the Optimal Physical Monopole for systems with non-zero charge 
and the Optimal Physical Dipole for systems with zero charge but non-zero dipole moment. 
These 
analytic expressions not only provide physical insight but are 
more computationally efficient than the numerical minimization procedures that 
are in general required to obtain an approximate charge distribution. Thus,
these analytic expressions may be particularly useful in applications such as molecular
dynamics where computational speed is critical. 

\subsection{Optimal Physical Monopole}

The Optimal Physical Monopole, order $k=0$, consists of a single charge. 
As long as the single charge has magnitude $\bar{q}=q$, i.e. is equal to the total 
charge of the original distribution,
the monopole term of the error expansion will obviously be zero. Thus,
any charge with magnitude equal to the net charge of the original distribution
will satisfy the first condition
which defines a Optimal Physical Multipole. Now, 
the remaining parameters, namely the position of the charge, are
chosen to satisfy condition 2 of the Optimal Physical Multipole definition,
namely to minimize the dipole term in the error expansion. 
In this particular case, the  $k+1$ term of the error, i.e. the dipole term,
can be made identically zero by solving 
\begin{eqnarray}
   p_x - \bar{q} \cdot x & = & 0 \\
   p_y - \bar{q} \cdot y & = & 0 \\
   p_z - \bar{q} \cdot z & = & 0
\end{eqnarray}
for $x,y,z$ where 
$p_x, p_y, p_z$ are the $x, y, z$ components
of the dipole moment $\boldsymbol{p}$ of the original distribution. 
Solving the above equations we have
\begin{eqnarray}
   \bar{q} & = & q \\
   \bar{\boldsymbol{r}} & = & \frac{\boldsymbol{p}}{q}
\end{eqnarray}
So, a charge of magnitude $\bar{q}$ placed at $\boldsymbol{\bar{r}}$ 
satisfies both conditions which define Optimal Physical Multipole of order $k=0$.
\begin{figure}[h!]
\centering
   \includegraphics[width=10cm]{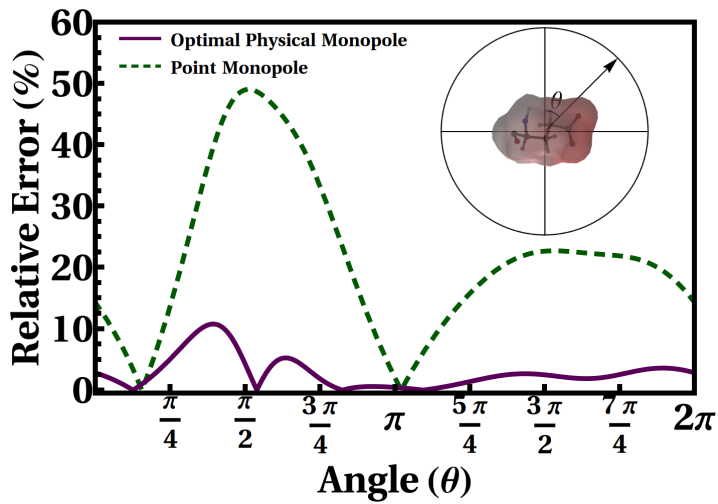}
   \caption{
Relative error in 
electrostatic potential produced by an Optimal Physical Monopole 
representing the charge distribution of an amino
acid that carries a net charge of $+1|e|$ (glutamic acid at physiological
pH). The inset image shows the plane bisecting the molecule which defines the angle 
plotted in the figure; to illustrate the original charge distribution, the 
electrostatic potential of the glutamic acid is plotted on its molecular surface as a color map
with blue representing positive and red representing negative potential \cite{Gordon2008}. 
The relative error is computed as in Fig. 1, (ratio of the absolute error to the RMS average
of the reference potential), at a distance $2 R_{0}$ which as before is twice the distance from the
center of geometry to the outermost charge. 
The Optimal Physical Monopole, located at the center of charge, produces 
less than 11\% error (solid purple line) for all angles
whereas the point monopole, located at the center of geometry,
produces over 45\% error (dashed green line) for some angles.
} 
\label{fig:Physical_Monopole}

\end{figure}

The Optimal Physical Monopole can offer substantial improvements
compared to 
the use of a single charge placed at the common choice of the geometric
center. In particular, there are biologically relevant examples, see
Fig. (\ref{fig:Physical_Monopole}), where a point monopole at the
origin produces an RMS error 5 times greater than that of the proposed
Optimal Physical Monopole.

Despite its similarity to the center of mass, the existence of negative
charge makes the center of charge term fundamentally different from
center of mass. In particular, as the total charge in the distribution
approaches zero, the center of charge tends towards infinity. 
However, in practice, charge is discrete, so there is a limit to the
maximum distance the center of charge can fall outside of a given
distribution. 

\subsection{Optimal Physical Dipole}

For an uncharged distribution, the Optimal Physical Dipole 
consists of two charges $\bar{q}_{1}=\bar{q}$
and $\bar{q}_{2}=-\bar{q}$
located at positions $\boldsymbol{\bar{r}_{1}}$ 
and $\boldsymbol{\bar{r}_{2}}$ respectively. Thus, it takes
7 parameters to uniquely define an Optimal Physical Dipole. By setting
\begin{equation}
   \bar{q}({\boldsymbol{\bar{r}_{1}}}-\\
   \boldsymbol{\bar{r}_{2}}) = \sum_{n=1}^{N}q_{n}\boldsymbol{r_{n}}
   \label{eq:DIPOLE_CONDITION}
\end{equation}
the dipole term in the error is zero 
and the two charges satisfy condition 1 of 
the Optimal Physical Dipole definition. Now, we will
rewrite the positions ${\boldsymbol{\bar{r}_{1}}}$ 
and $\boldsymbol{\bar{r}_{2}}$ in the following form:
\begin{eqnarray}
   \boldsymbol{\bar{r}_{1}} & = & \boldsymbol{\bar{d}}+\frac{\sum q_{n}\boldsymbol{r_{n}}}{2\bar{q}}\nonumber \\
   \boldsymbol{\bar{r}_{2}} & = & \boldsymbol{\bar{d}}-\frac{\sum q_{n}\boldsymbol{r_{n}}}{2\bar{q}}
\label{eq:phys_dip_pos}
\end{eqnarray}
where $\boldsymbol{\bar{d}}$ represents the geometric center between
the two charges. We can
see that these positions satisfy relation (\ref{eq:DIPOLE_CONDITION})
automatically. By writing the positions of the charges in this manner,
we have divided the process
of determining the remaining parameters which define 
the Optimal Physical Dipole into two steps, namely, finding the 
optimal placement of the charges, $\boldsymbol{\bar{d}}$, and
finding the optimal magnitude of the charge, $\bar{q}$.
Note that finding the optimal charge value fixes the separation
between the two charges, since the dipole moment of 
the representative distribution has been constrained to equal the original
dipole moment. 
\begin{figure}[h!]
\centering
   \includegraphics[width=10cm]{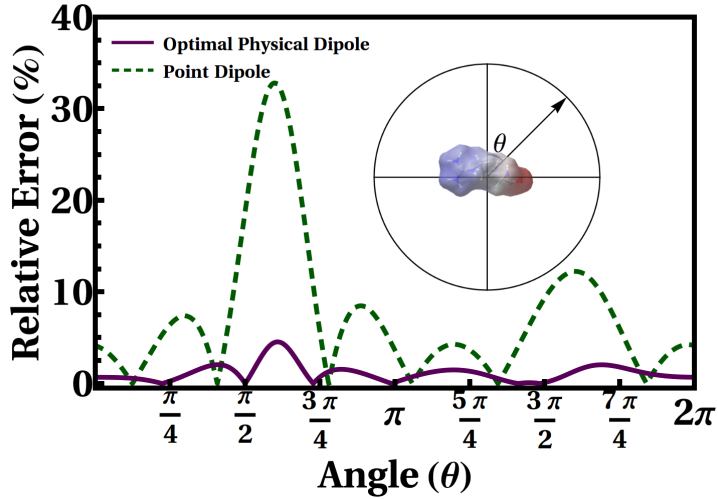}
\caption{\label{fig:Physical-versus-Point}
Relative error in
electrostatic potential produced by Optimal Physical Dipole
representing charge distribution of an amino
acid that carries a zero net charge (C-terminal arginine at physiological
pH). The inset image shows the plane bisecting the molecule defining the angle 
plotted in the figure; to illustrate the original charge distribution, the 
electrostatic potential of the arginine is plotted on its molecular surface as a color map
with blue representing positive and red representing negative potential \cite{Gordon2008}.
The relative error is computed as in Fig. 1, (ratio of the absolute error to the RMS average
of the reference potential), at a distance $2 R_{0}$, which as before is twice the distance from the
center of geometry to the outermost charge. 
The Optimal Physical Dipole produces 
less than 5\% error (solid purple line) for all angles,
whereas the optimally placed, see \cite{Gramada2008,Platt1996}, point dipole
produces over 30\% error (dashed green line) for some angles.
}
\end{figure}

The placement of the geometric center of the charges
composing the 
Optimal Physical Dipole, $\boldsymbol{\bar{d}}$, that minimizes the quadrupole term of the 
error expansion is given by 
\begin{equation}
\bar{d}_{k}=\frac{2}{3 p^{2}}
  \left(\sum_{i=x,y,z} Q_{ki} p_{i}-
  \left(\frac{\sum_{i,j=x,y,z} Q_{ji} p_{i} p_{j}}
  {4 p^{2}}\right)p_{k}\right)
\label{eq:center_of_dipole}
\end{equation}
where $k=x,y,z$ and thus the $\bar{d}_{k}$'s are the components of $\boldsymbol{\bar{d}}$. 
This optimal position, known as the center of dipole, 
was derived previously \cite{Platt1996} for a different purpose, namely for 
matching point multipole expansions between different charge
distributions. Now, unlike the point dipole,
the Optimal Physical Dipole has physical size and thus an additional parameter
with which to further minimize the error with respect to the given
potential. In other words, Eqs. (\ref{eq:center_of_dipole}) and (\ref{eq:DIPOLE_CONDITION}),
determine only 6 of the 7 parameters required to define an Optimal Physical Dipole.
Since the quadrupole moment is the lowest order non-zero
term remaining in the error expansion, by choosing the optimal charge
value, we want to further minimize the quadrupole term in the error.
However, for any charge value $\bar{q}$ an Optimal Physical Dipole 
placed at the center of dipole has no quadrupole
moment as can be seen by setting $N=2$, substituting the center of dipole,
Eq. (\ref{eq:center_of_dipole}) and 
$q_{1}=-q_{2}=\bar{q}$ into Eq. (\ref{eq:phys_dip_pos}) then substituting these
variables into Eq. (\ref{eq:multipoles}). Thus, the
quadrupole term in the error, Eq. (\ref{eq:MULTIPOLE_RMS_ERROR_SIMP})
is unaffected by the choice of the charge magnitude, $\bar{q}$ and the
quadrupole term has already been globally minimized. 
Therefore, to uniquely define
the charge, $\bar{q}$, we follow the OPM definition and globally minimize
the next term in the error expansion, namely the
octupole term. Specifically, if we consider the $\ell=3$ term of
Eq. (\ref{eq:MULTIPOLE_RMS_ERROR_SIMP}), using the connection formula
from spherical multipoles to Cartesian multipoles we can 
compute
\begin{equation}
\sum_{i,j,k=x,y,z}\frac{\partial}{\partial \bar{q}}\left(O_{ijk}-\overline{O_{ijk}}\right)^{2}=0 \label{eq:min_oct}
\end{equation}
where $O_{ijk}$ and $\overline{O_{ijk}}$ are the octupole
moments, in Cartesian coordinates, of the original distribution and the Optimal 
Physical Dipole respectively,
for an expansion computed about the center of dipole.
Now, by noting that $\overline{O_{ijk}}$ is a function of
$\bar{q}$,  we find that
Eq. (\ref{eq:min_oct}) is satisfied when $\bar{q} \rightarrow \infty$ 
or if the
charge value is given by
\begin{equation}
\bar{q}=\sqrt{\frac{3 p^{6}}
{2\sum_{i,j,k=x,y,z} O_{ijk} p_{i} p_{j} p_{k}}}
\label{eq:optimal_charge}
\end{equation}
Thus, 
Eqs. (\ref{eq:phys_dip_pos}), (\ref{eq:center_of_dipole}) and (\ref{eq:optimal_charge})
define the Optimal Physical Dipole, i.e. defines the best placement of charges
such that RMS error on a sphere centered at the center of dipole is minimized.
For certain charge distributions relevant to molecular biophysics, 
the Optimal Physical Dipole produces error several times lower than
that of the corresponding a point dipole, see Fig. \ref{fig:Physical-versus-Point}.

Although the Optimal Physical Dipole can produce dramatically lower error
than the point dipole, it is possible that 
\begin{equation}
\left(\sum_{i,j,k=x,y,z} O_{ijk} p_{i} p_{j}p_{k}\right) \le 0
\label{eq:bad_octupole}
\end{equation}
In this case, the charge given by Eq. (\ref{eq:optimal_charge}) is imaginary.
This situation occurs when the orientation of the dipole with respect to the
octupole is such that increasing the distance between the charges
of the Optimal Physical Dipole always increases the error. Thus,
Eq. (\ref{eq:min_oct}) is formally satisfied only 
for $\bar{q}\rightarrow \infty$.
In a practical calculation, a physical dipole with the above
property is constructed by fixing the
separation between the charges $||\boldsymbol{\bar{r}_{1}}-\boldsymbol{\bar{r}_{2}}||$ 
to a very small value, while increasing
the charge accordingly to maintain the original dipole moment. 
In these cases,
the Optimal Physical Dipole does not offer an advantage over the optimal 
point dipole, however,
the Optimal Physical Dipole can always mimic the point dipole to arbitrary
precision and thus the two distributions will produce equivalent error. 
Thus, even if inequality (\ref{eq:bad_octupole}) holds, the Optimal Physical
Dipole represents the optimal placement of two point charges
and is at least as accurate as the point dipole. 

\begin{figure}[h!]
\centering
\includegraphics[width=10cm]{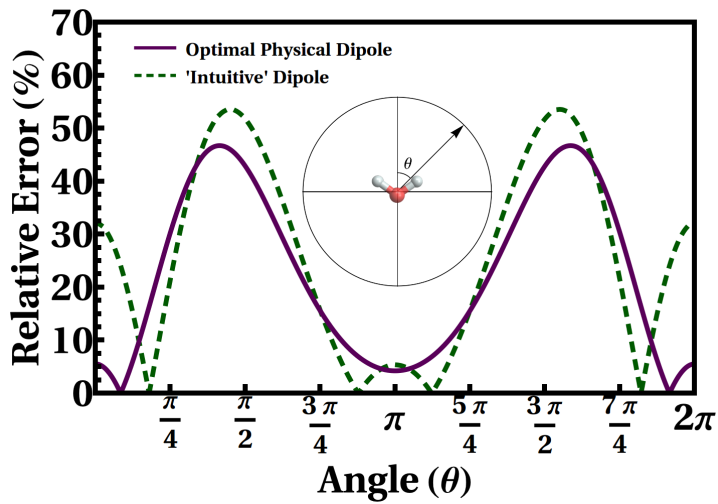}
\caption{\label{fig:water}
Relative error in computed electrostatic potential for a
water molecule: Optimal vs. `intuitive' Physical Dipole. 
The reference charge density of the electron distribution
for water was calculated using the CCSD method
with aug-cc-pCVTZ basis set \cite{Dunning:89,Kendall92,Woon:95}
at experimental equilibrium geometry of water in vacuum. The electron 
distribution is confined to a box with a side length of $4 a_{0}$;
the discretization is $0.1 a_{0}$ along each axis, 
where $a_{0}$ is the atomic unit
of length (the Bohr Radius). 
Here, the Optimal Physical Dipole consists of a negative
and positive charge with magnitude $10 |e|$ whose center
is offset by $0.070 a_{0}$ from the oxygen nucleus toward 
the protons along the water symmetry axis. The charges 
are separated by $0.074 a_{0}$ to maintain the original dipole
moment of water. 
The `intuitive' dipole representation,
places a positive charge, with magnitude $0.7 |e|$, 
at the geometric center
between the protons and a negative charge ($-0.7 |e|$) on the oxygen.
The relative error is computed as in Fig. 1, (ratio of the absolute error to the RMS average
of the reference potential), at a distance $2 R_{0}$ which in this case has been defined to be
5 times the OH distance from the center of the quantum electron charge distribution.
Along some directions, e.g. at $\theta=0=2\pi$, 
the Optimal Physical Dipole produces error 6 times smaller
than that of the `intuitive' dipole. Overall, the average RMS error of the
potential due to the Optimal Physical Dipole is 10\% smaller than that 
due to the `intuitive' dipole.} 
\end{figure}

Curiously, the most
biologically important molecule, water, satisfies inequality 
(\ref{eq:bad_octupole}).
Thus, among point charge representations that use only two charges to 
approximate 
the true charge density distribution 
of water molecule, 
the most accurate electrostatic potential outside of the original charge 
distribution is produced by
two charges that mimic a point dipole 
placed at the center of dipole, that is itself a special case 
of the Optimal Physical Dipole with $||\boldsymbol{\bar{r}_{1}}-\boldsymbol{\bar{r}_{2}}|| \rightarrow 0$. 
Other two-charge charge models that may appear more `intuitive' 
by keeping the charge-charge distance
comparable to the size of the system, i.e. water, 
are bound to produce less accurate representations of the electrostatic
potential of the original charge distribution. For example, the 
model used in \cite{Izvekov2005},
places a $-0.7 |e|$ charge on the oxygen and $+0.7 |e|$ at the geometric
center of the two hydrogens. However,
this `intuitive' dipole solution produces 20\% more error at positions near the
hydrogen atoms and overall produces $\sim10\%$ higher RMS error compared to 
Optimal Physical Dipole solution (when integrated over a sphere
of radius $2R_{0}$), see Fig. \ref{fig:water}.  We stress that the above 
accuracy analysis is strictly limited to electrostatics; other considerations
may dictate electrostatically suboptimal placements of point charges in 
coarse-grained models of water.  

\section{Cost Benefit Analysis}

Whether the OPMs introduced in the preceding sections
will be a beneficial simplification in practice 
will depend upon the specifics of the
application. Physical multipoles may be easier to 
implement than point multipoles
in applications that already utilize point charges, such as many 
molecular dynamics codes. Furthermore, the lowest order Optimal Physical
Multipoles have been shown to be always at least as
accurate as the corresponding point multipole and in certain distributions
may be substantially more accurate than an equivalent point multipole.
We conjecture that the accuracy advantage for some practically 
useful distributions also holds for higher order OPMs. 

One of the major advantages in using the 
Optimal Physical Multipole method versus
point multipoles comes in at the level of pairwise interactions. Note that
although Cartesian formalisms exist for the calculation of pairwise interactions \cite{Kong1997},
we will restrict the following discussion to the more commonly used spherical formalism.
Consider two systems of charges, one containing of $N_{A}$ 
charges $\{q_{i}\}$ located at 
$\{\boldsymbol{r_{i}}\}$ clustered around an origin located
at $\boldsymbol{R_{A}}$ and another containing of $N_{B}$ 
charges $\{q_{j}\}$ located at 
$\{\boldsymbol{r_{j}}\}$ clustered around an origin located
at $\boldsymbol{R_{B}}$. Then the interaction energy between the
two systems is given by
\begin{equation}
U_{AB}=\sum_{i=1}^{N_{A}}\sum_{j=1}^{N_{B}}
\frac{q_{i}q_{j}}{4\pi\epsilon_{0}||\boldsymbol{r_{i}}-\boldsymbol{r_{j}}||}
\end{equation}
Thus, the interaction requires $N_{A}N_{B}$ computations. So, for
large charge distributions the computation scales as $O(N^{2})$.
Now, using the point multipole expansion, this interaction energy becomes
\cite{Stone1984}
\begin{eqnarray}
U_{AB} & = & \frac{1}{\epsilon_{0}}\sum_{\ell_{A}=0}^{\infty}\sum_{\ell_{B}=0}^{\infty}(-1)^{\ell_{B}}\left(\begin{array}{c}
2\ell_{A}+2\ell_{B}\\
2\ell_{A}\end{array}\right)^{1/2} \frac{1}{\sqrt{(2\ell_{A}+1)(2\ell_{B}+1)}}\nonumber \\
 &  & \times\sum_{m_{A}=-\ell_{A}}^{\ell_{A}}\sum_{m_{B}=-\ell_{B}}^{\ell_{B}}\Biggl(I_{\ell_{A}+\ell_{B}}^{-m_{A}-m_{B}}(\boldsymbol{R_{AB}})\nonumber \\
 &  & q_{\ell_{A}}^{-m_{A}}q_{\ell_{B}}^{-m_{B}}\langle\ell_{A},m_{A};\ell_{B},m_{B}|\ell_{A}+\ell_{B},m_{A}+m_{B}\rangle\Biggr)
\end{eqnarray}
for the irregular spherical harmonic, 
$I_{\ell_{A}+\ell_{B}}^{-m_{A}-m_{B}}(R_{AB})$, given by 
\begin{equation}
I_{\ell_{A}+\ell_{B}}^{-m_{A}-m_{B}}(\boldsymbol{R_{AB}})=
\left(\frac{4\pi}{2\ell_{A}+2\ell_{B}+1}\right)^{1/2}
\frac{Y_{\ell_{A}+\ell_{B}}^{-(m_{A}+m_{B})}(\boldsymbol{\hat{R}_{AB}})}{R_{AB}^{\ell_{A}+\ell_{B}+1}}.
\end{equation}
where 
$\boldsymbol{R_{AB}}=||\boldsymbol{R_{A}}-\boldsymbol{R_{B}}||$, $\boldsymbol{\hat{R}_{AB}}$ is the unit
vector in the direction of $\boldsymbol{R_{AB}}$ 
and the bracketed expression is the Clebsch-Gordan Coefficient. This
expression has the benefit of scaling linearly with the number of
charges. However, the number of computations required to compute
the pairwise interaction energy between two distributions
scales as $O(\ell^{3})$ with the number, $\ell$ of point multipole terms in 
the interaction. In particular,
to determine the interaction energy between two charge distributions
each with zero net charge but non-zero dipole moment requires the calculation
of 9 terms, namely the dipole-dipole interaction term. On the other
hand, an equivalent Optimal Physical Dipole representation requires only 4
calculations. To determine the interaction energy of a system with
non-zero monopole, dipole and quadrupole moments, 
a scheme based on point multipoles would
require, in general, 84 terms, namely the various monopole-monopole,
monopole-dipole, monopole-quadrupole, dipole-dipole, dipole-quadrupole
and quadrupole-quadrupole terms. On the other hand, the interaction 
energy of two Optimal Physical Quadrupoles, 
containing 5 charges each, requires only 25 terms. Additionally, the Coulomb
interaction, used to compute the interaction between Optimal Physical Multipoles, does not
require the computation of spherical harmonics or Clebsch-Gordan coefficients, 
which may be cumbersome. 
Thus, using Optimal Physical Multipoles has a computational advantage over 
spherical point multipoles when calculating interaction energies. We have
kept the above discussion at the most general level -- specific symmetries 
present in the original charge distribution may change the numbers. 

\section{Conclusion}

Point multipole expansion is a widely used approach to 
simplify and approximate potentials and fields around complex charge 
distributions via power 
series. Hoever, in practical calculations which typically retain only a few
leading terms, the approximation can lead to 
inaccurate results in the near-field.
In this work, we have introduced 
an alternative to the point multipole expansion for
approximating the electrostatic potential of a charge distribution, namely,
Optimal Physical Multipoles (OPMs).
An OPM consists of a set of point charges which
are optimally placed to  best reproduce the original charge distribution
at a given multiople order. Namely, the OPM 
exactly reproduces the point multipole expansion of the original distribution up
to a given order while optimally approximating the remaining lowest order non-zero point multipole terms.
Thus, by construction, OPMs retain many of the useful properties of point
multipole expansions, in particular they retain the asymptotic behavior of the
point multipole expansion. At the same time, an expansion based on 
OPMs can be more accurate
than the point multipole expansion of the same order; for some 
charge distributions relevant to molecular biophysics (amin-acid) the 
accuracy gain is more than 5-fold in the near-field. At least 
for the lowest order OPMs (monopole and dipole), the expansion is 
guaranteed to be at least as accurate as the corresponding 
point multipole expansion of the same order.  We have provided
a general framework for calculating OPMs to any order and have derived
closed-form expressions for the Optimal Physical Monopole and Dipole. 

In comparison to point multipoles, 
expansions based on Optimal Physical Multipoles have many 
desirable properties that may be useful
in practical computations; in particular, their mathematical form -- 
the sum of contribution from point sources -- is simpler that that 
of the conventional point multipole expansion. Thus, OPMs 
may be easier to implement into existing
molecular dynamics protocols. At the same time, unlike many approximations 
based on point charge representations, OPMs are uniquely defined and 
preserve the natural hierarchy of the multipole features of 
the original charge distribution. 

	Optimal Physical Multipoles is a new concept; thus its many applications and potentially useful properties
remain unexplored in this proof-of-concept work. OPMs are expected to have 
utility in coarse-grained \cite{Basdevant2007,Izvekov2005a} 
and multi-scale methods \cite{Anandakrishnan2010},
especially in dynamics \cite{Anandakrishnan2011} 
where analytic expressions and the simplicity of the 
algorithms is key. The 
Optimal Physical Multipole framework 
provides a systematic way of deriving approximate charge distributions that have the potential
to be both computationally effective and produce an accurate representation of the original electrostatic potential.
To further improve the representation of the original potential via 
OPMs, future
work may consider partitioning the original charge distribution into 
several domains, and finding OPMs for each of them separately, 
similar to the distributed multipoles approach \cite{Gramada2011,Stone1981,Stone1985}. Further exploration of the mathematical and physical 
properties of OPMs is also desirable; 
future studies may reveal whether closed form
solutions exist for higher order OPMs.


\section{Acknowledgements}

The authors wish to thank Edward Valeev for providing the 
high quality electron density distribution for water.
Support from the NIH 2R01 GM076121 is acknowledged.
\bibliographystyle{plain}
\bibliography{physical_multipoles}







\end{document}